\documentclass[usegraphicx,useAMS,usenatbib]{mn2e}
\usepackage{amsmath}

\usepackage{times}

\newcommand{\apj}{ApJ}

\newcommand{\mnras}{MNRAS}

\title[V1154~Cyg revisited]{The {\boldmath $Kepler$} Cepheid V1154~Cyg revisited: light curve modulation and detection of granulation}

\author[A. Derekas et~al.]{A. Derekas$^{1,2}$\thanks{E-mail:
derekas@gothard.hu}, E. Plachy$^{2}$,  L. Moln\'ar$^{2}$, \'A. S\'odor$^{2}$, J. M. Benk\H o$^{2}$, L. Szabados$^{2}$,   
\newauthor Zs. Bogn\'ar$^{2}$, B. Cs\'ak$^{1}$, Gy. M. Szab\'o$^{1,2}$, R. Szab\'o$^{2}$, A. P\'al$^{2,3}$\\
\\$^1$ELTE Gothard Astrophysical Observatory, H-9704 Szombathely, Szent Imre herceg \'ut 112, Hungary
\\$^2$Konkoly Observatory, Research Centre for Astronomy and Earth Sciences, Hungarian Academy of Sciences, H-1121 Budapest, \\
Konkoly Thege Mikl\'os \'ut 15-17, Hungary
\\$^3$Department of Astronomy, E\"otv\"os Lor\'and University, P\'azm\'any P\'eter s\'et\'any 1/A, H-1117 Budapest, Hungary\\
}

\begin{document}

\date{Accepted ... Received ..; in original form ..}


\maketitle

\begin{abstract}

We present a detailed analysis of the bright Cepheid-type variable star V1154~Cygni using 4 years of continuous observations by the {\it Kepler} space telescope. We detected 28 frequencies using standard Fourier transform method. We identified modulation of the main pulsation frequency and its harmonics with a period of $\sim$159~d. This modulation is also present in the Fourier parameters of the light curve and the O--C diagram. We detected another modulation with a period of about 1160~d. The star also shows significant power in the low-frequency region that we identified as granulation noise. The effective timescale of the granulation agrees with the extrapolated scalings of red giant stars. Non-detection of solar-like oscillations indicates that the pulsation inhibits other oscillations. We obtained new radial velocity observations which are in a perfect agreement with previous years data, suggesting that there is no high mass star companion of V1154~Cygni. Finally, we discuss the possible origin of the detected frequency modulations.

\end{abstract}


\section{Introduction}

There is only one genuine Cepheid variable in {\it Kepler} field: V1154~Cyg (KIC~7548061). It has a mean V brightness of $\sim$9.1~mag and a pulsation period of $\sim$4.925~d. Its brightness variation was discovered by \citet{str63}. Further multicolor photoelectric and CCD photometric observations were published by \citet{wac76,sza77,are98,ign00,ber08,pig09}, while radial velocity data were published by \citet{gor98,imb99}. Detailed spectroscopic analysis was performed by \citet{luc06} and \citet{mol08} who determined the basic atmospheric parameters. \citet{jer13} has derived the radius and distance using Baade-Wesselink technique as ${\rm 44.5 \pm 4.1~R_{\odot}}$ and ${\rm 2100 \pm 200~pc} $.

Although V1154~Cyg was already observed from space by the {\it Hipparcos} satellite \citep{esa97} and OMC onboard {\it INTEGRAL}, the accuracy of these observations is behind compared to the {\it Kepler} photometry. The first results based on the first four quarters of {\it Kepler} data were published by \citet{sza11} who found V1154~Cyg to be single periodic. The frequency analysis did not reveal any additional pulsation frequency while the period remained stable during the last 40 years.

\citet{der12} analysed 600 days of {\it Kepler} data. The data revealed cycle-to-cycle fluctuations in the pulsation period, indicating that classical Cepheids may not be as accurate astrophysical clocks as commonly believed. A very slight correlation between the individual Fourier parameters and the O--C values was found, suggesting that the O--C variations might be due to the instability of the light-curve
shape. This period jitter in V1154~Cyg represents a serious limitation in the search for binary companions as the astrophysical noise can easily hide the signal of the light-time effect.

\begin{figure*}
\begin{center}
\includegraphics[width=17cm]{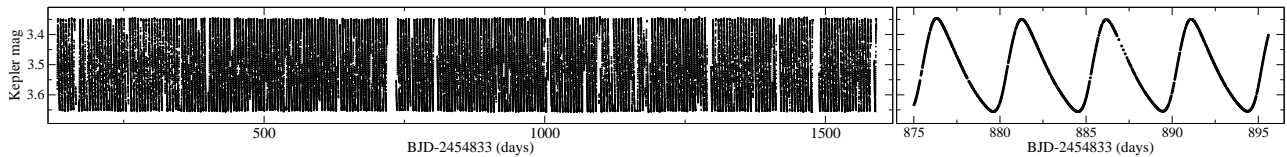}
\end{center}
\caption{\label{lc} {\it Left panel:} Detrended long cadence {\it Kepler} observations of V1154~Cyg. {\it Right panel:} Zoom in four cycles of long cadence data.}
\end{figure*}

Later, \citet{eva15} also studied the light curve stability of RT Aur, a fundamental mode and SZ~Tau, an overtone mode Cepheid based on {\it MOST} data. They found cycle-to-cycle light curve variation of SZ~Tau and argued that it is the instability in the pulsation cycle and also a characteristic of the O--C curves of overtone pulsators whose oscillation seems to be less stable than that of the fundamental mode pulsator at both long and short timescales.

The latest study of V1154~Cyg was performed by \citet{kanev} using the 4 years of long cadence {\it Kepler} data and studied the light curve by using Fourier decomposition technique. They found that the Fourier parameters $R_{21}$ and $R_{31}$ (see in Sect.\ \ref{foupars}) show modulation with a period of 158.2~d. They concluded that this modulation of the light curve is very similar to the phenomenon of the Blazhko effect in RR~Lyrae stars.

In this paper, we present the analysis of V1154~Cyg using the whole {\it Kepler} dataset spanning 1460 d of continuous observations. In Sect.\ \ref{data}, we briefly describe the data used and details of the pixel-level photometry. In Section\ \ref{lcsection} we present the results of the Fourier analysis and the Fourier decomposition of the light curve, while the construction of the  O--C diagram is shown Section~\ref{oc}. We combined our new radial velocities (RVs) and previous RV data in Section~\ref{newrv}.  In Section~\ref{discussion} we discuss the detection of granulation noise, its properties and the origin of the frequency modulation. In Sect.~\ref{summary} we summarize our results.

\section{Kepler observations}
\label{data}

The photometric data we use in our analysis were obtained by the {\it Kepler} space telescope. The telescope was launched in March 2009 and designed to detect transits of Earth-like planets. A detailed technical description of the {\it Kepler} mission can be found in \citet{koc10} and \citet{jen10a,jen10b}. 

The {\it Kepler} space telescope originally observed a 105 square degree area of the sky in the constellations Cygnus and Lyra, and has two observational modes, sampling data either in every 58.9~s (short-cadence whose characteristics was analysed by \citealt{gil10}) or 29.4~min (long-cadence, hereafter LC), providing quasi-continuous time series for hundreds of thousands of stars. V1154~Cyg was observed in LC mode during the entire mission (Q0--Q17) spanning 1470 days. Data obtained during Q0 (covering 10 days altogether) have been omitted because an unidentified timing error would have falsified the results, truncating the useful data set to 1460 d. The star was also observed in short cadence (SC) mode in 8 quarters (Q1, Q5, Q6, Q13--Q17).

We note that we use magnitude scale in the analyses in Sections\ \ref{lcsection} and \ \ref{oc} because it is commonly used in the study of the classical variables like Cepheids. However, we use flux scale in Section~\ref{discussion} because it is generally used in studying solar-like oscillations.

\subsection{Cepheid pixel photometry}

We applied our ``tailor-made'' aperture photometry to the 4 year long {\it Kepler} pixel data. This method already proved to be useful in the {\it Kepler} RR~Lyrae data processing \citep{ben14,ben15} and it is described in detail in those works. Briefly, we check the flux variations for the individual pixels separately, and include all that contain the stellar signal in the aperture for a given quarter. 
The apertures defined by the pixels that clearly contain the flux variation of the targets are typically larger than the pre-defined optimal apertures, and provide more precise light curves. Thus we performed a detailed investigation of the flux variation curves in the individual pixels of the target pixel data and we found that the variability of V1154~Cyg is dominant in all pixels in the pixel mask in each quarter. This implies a significant flux loss due to the tight mask, similar to the case of some RR~Lyrae stars \citep{ben14}.

The target pixel mask contains two other stars, one of which falls exactly in the charge blooming columns of V1154~Cyg, therefore it is not possible to avoid contaminations. However, the other star, 2MASS J19481541+4307203, which is a variable, can be separated. We examined the effect of separating this second star carefully. We defined a tailor-made aperture for this star too, and we found that within this aperture the flux from V1154~Cyg is a factor of ten larger than that of the examined star itself. We concluded that the removal of pixels containing 2MASS J19481541+4307203 causes larger uncertainties in the data than its contamination. Furthermore, the few mmag-level light variations of this star do not affect our analysis, and cannot be responsible for any modulation detected in the light curve of V1154~Cyg. Therefore, we decided to include these pixels in the total aperture and use all the pixels of the target pixel mask.

Our quarter stitching and scaling method was identical to the technique of \citet{ben14}. The last step before the analysis was the de-trending process. Instead of using a polynomial fit or moving average technique we found an alternative method to be more suitable. We calculated the mean of the upper and lower envelopes of the light curve and then subtracted it. A sample of the final light curve is presented in Fig.\ \ref{lc}.

\section{Light curve analysis}
\label{lcsection}

Having four years of continuous photometric data, it is possible to study the long term stability of the light curve. First, we performed the frequency analysis of the dataset, then we calculated the Fourier parameters of each pulsation cycle to study changes in the light curve shape.

\subsection{Frequency analysis}
\label{fouanal}

The frequency content of the light curve of V1154~Cyg was investigated with the standard Fourier transform method by using {\sc Period04}  \citep{len05}. Least-squares fitting of the parameters was also included and the signal-to-noise ratio ($S/N$) of each frequency was calculated following the method of \citet{bre93}. 

\begin{figure}
\begin{center}
\includegraphics[width=8.5cm]{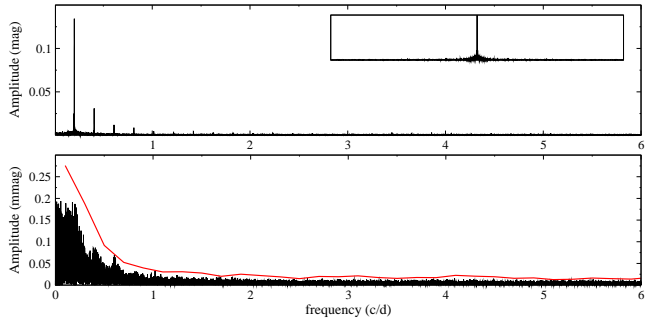}
\end{center}
\caption{\label{foursp} {\it Top panel:} Frequency spectrum of V1154~Cyg based on Q1-Q17 LC data. The insert shows the window function of the data. {\it Bottom panel:} The residual frequency spectrum after subtracting 28 frequencies. The red line shows the significance limit. Note the change in the scale (i.e.\ mag vs. millimag) on the vertical axis.}
\end{figure}

We detected 28 significant frequencies altogether, following the recipe of \citet{bre93}. The frequencies, amplitudes and phases are listed in Table\ \ref{fourtbl}, while the frequency spectrum is shown in Fig.\ \ref{foursp}. The most striking peaks in the frequency spectrum are the main pulsation frequency at $f_{1}=0.2030246$ c/d and its harmonics up to the 11th order with very low amplitudes. 

We identified modulation of the main pulsation frequency and its second and third harmonics (see last column in Table\ \ref{fourtbl}) with a modulation frequency of $f_{\rm m1}=0.0063$~c/d (corresponding to $\sim$ 159~d). This periodic modulation was also detected in the Fourier parameters (Sect.\ \ref{foupars}) and in the O--C diagram (Sect.\ \ref{oc}). 

We also detected another modulation of the main pulsation frequency with $f_{\rm m2}=0.00086$~c/d ($\sim$ 1160~d). We note that this modulation is relatively close to the length of the whole dataset, but it is definitely shorter by 300~days. Future observations may reveal its true nature.

\setlength{\tabcolsep}{4pt}
\begin{table}   
\begin{center}
 \caption{Results of the frequency analysis of V1154~Cyg based on Q1-Q17 LC {\it Kepler} data. $f_{1}$ is the main pulsation frequency. Besides identifying the harmonics of $f_{1}$, we also detected the modulation of $f_{1}$, $f_{2}$, $f_{3}$  with a period of 159~d and 1160~d, respectively.}  
\label{fourtbl} 
\begin{tabular}{ccccc}
\hline   
No. & Frequency & Amplitude & Phase & Identification \\
    &     (d$^{-1}$)      &  (mmag)    &   (rad/2$\pi$)          &        \\
\hline  
$f_{1}$    &	0.203027027(24)  &  139.662(14)   &  0.762573(10)   &          \\
$f_{2}$    &	0.40605405(9)  &  37.925(14)   &  0.973077(37)   &	2$f_{1}$   \\
$f_{3}$    &	0.60908108(35)   &  9.583(14)   &  0.21328(15)   &	3$f_{1}$   \\
$f_{4}$    &	0.8121081(31)    &  1.066(12)   &  0.4629(13)   &	4$f_{1}$   \\
$f_{5}$    &	1.015135(6)    &  0.594(14)   &  0.0687(24)   & 	5$f_{1}$   \\
$f_{6}$    &	1.218162(6)    &  0.591(13)   &  0.3080(24)   &	6$f_{1}$   \\
$f_{7}$    &	1.421189(9)    &  0.374(14)   &  0.520(4)   &	7$f_{1}$   \\
$f_{8}$    &	1.624216(16)   &  0.209(14)   &  0.739(7)   &	8$f_{1}$   \\
$f_{9}$    &	1.827243(29)   &  0.117(13)   &  0.951(12)   &	9$f_{1}$   \\
$f_{10}$   &	2.03027(6)   &  0.055(14)   &  0.177(25)   &	10$f_{1}$   \\
$f_{11}$   &	0.006327(9)    &  0.371(15)   &  0.8404(38)   &   $f_{\rm m1}$       \\
$f_{12}$   &	0.196700(6)    &  0.554(14)   &  0.3927(25)   &	$f_{1}-f_{\rm m1}$  \\
$f_{13}$   &	0.209354(10)   &  0.319(13)   &  0.884(4)   &	$f_{1}+f_{\rm m1}$   \\
$f_{14}$   &	0.399727(12)   &  0.289(13)   &  0.346(5)   &	$f_{2}-f_{\rm m1}$   \\
$f_{15}$   &	0.602754(30)   &  0.111(15)   &  0.488(13)   &	$f_{3}-f_{\rm m1}$   \\
$f_{16}$   &	0.81757(8)   &  0.042(15)   &  0.037(34)   & $f_{4}$+$f_{\rm m1}$--$f_{\rm m2}$   \\
$f_{17}$   &	1.02060(12)  &  0.028(13)   &  0.252(5)   &	$f_{5}$+$f_{\rm m1}$--$f_{\rm m2}$   \\
$f_{18}$   &	0.412381(28)   &  0.121(14)   &  0.729(11)   &	$f_{2}+f_{\rm m1}$   \\
$f_{19}$   &	0.190374(15)   &  0.219(13)   &  0.049(6)   & $f_{1}-{\rm 2}f_{\rm m1}$   \\
$f_{20}$   &	0.393401(19)   &  0.171(13)   &  0.984(8)   &	$f_{2}-{\rm 2}f_{\rm m1}$   \\
$f_{21}$   &	0.59643(5)   &  0.069(15)   &  0.178(20)   &	$f_{3}-{\rm 2}f_{\rm m1}$   \\
$f_{22}$   &	0.00086(5)   &  0.068(14)   &  0.068(20)   &    $f_{\rm m2}$	\\
$f_{23}$   &	0.202163(8)    &  0.413(15)   &  0.6813(34)   &	$f_{1}-f_{\rm m2}$   \\
$f_{24}$   &	0.203891(5)    &  0.606(15)   &  0.7655(23)   &	$f_{1}+f_{\rm m2}$   \\
$f_{25}$   &	0.201300(8)    &  0.430(13)   &  0.4778(33)   &	$f_{1}-{\rm 2}f_{\rm m2}$   \\
$f_{26}$   &	0.204754(7)    &  0.458(14)   &  0.0240(30)   & 	$f_{1}+{\rm 2}f_{\rm m2}$   \\
$f_{27}$   &	2.23330(10)  &  0.033(13)   &  0.40(4)   &	11$f_{1}$   \\
$f_{28}$   &	0.161610(13)   &  0.262(14)   &  0.598(5)   &           \\
\hline  
\end{tabular} 
\end{center}  
\end{table}
\setlength{\tabcolsep}{6pt}

\subsection{Fourier parameters}
\label{foupars}

\begin{figure}
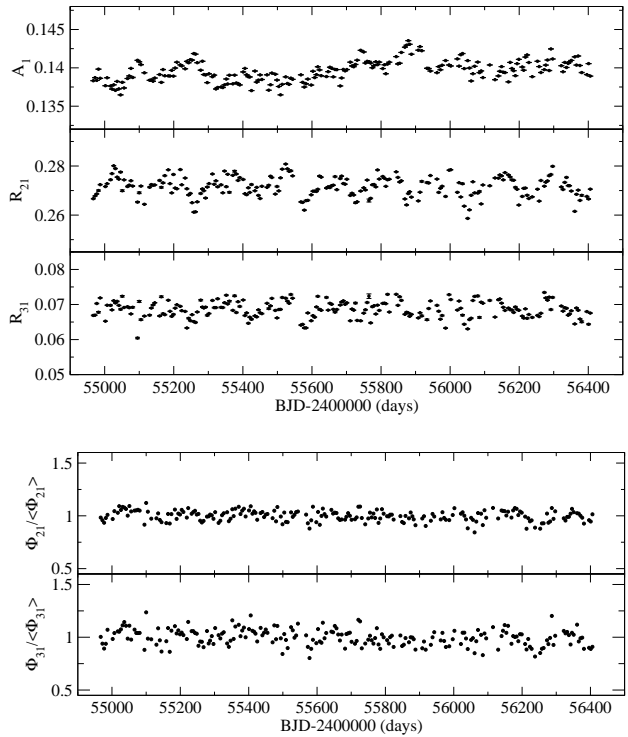

\begin{center}
\includegraphics[width=9.5cm]{fig3.eps}
\includegraphics[width=8cm]{fig4.eps}
\end{center}
\caption{\label{fourpar} {\it Top three panels:} The amplitude change and the variation of the $R_{21}$ and $R_{31}$. {\it Bottom two panels:} The relative variation of $\phi_{21}$ and  $\phi_{31}$.}
\end{figure}

In order to study the change of the light curve shape we used the technique of Fourier decomposition and examined the temporal variation of the Fourier parameters. For this, we fitted an eighth-order Fourier polynomial at the primary frequency and its harmonics for each pulsation cycle: 
\begin{equation} m=A_{0}+\sum_{k=1}^{8} A_{k} \cdot \sin\left( 2 \pi ift + \phi_{i} \right), \end{equation} 
where {\it m} is the magnitude, {\it A} is the amplitude, {\it f} is the frequency, {\it t} is the time of the observation, $\phi$ is the phase and index {\it k} runs from 1 to 8. Then we characterised the light curve shapes with the Fourier parameters \citep{sim81}, of which we show particular results for $R_{21}=A_{2}/A_{1},\ R_{31}=A_{3}/A_{1}$, as well as $\,\phi_{21}=\phi_{2}- 2\phi_{1}$, and $\phi_{31}=\phi_{3}-3\phi_{1}$. The top three panels in Fig.\ \ref{fourpar} show the amplitude change and the variations of the $R_{21}$ and $R_{31}$ quantities, while the bottom two panels in Fig.\ \ref{fourpar} show the relative variations of $\phi_{21}$ and $\phi_{31}$.

\begin{figure}
\begin{center}
\includegraphics[width=8cm]{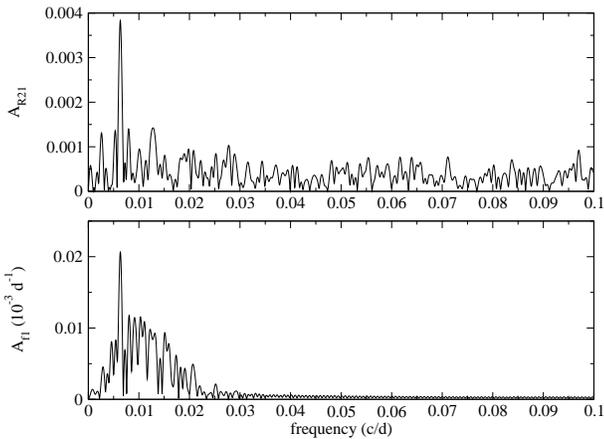}
\end{center}
\caption{\label{r21fou} {\it Top panel:}The frequency spectrum of the Fourier parameter $R_{21}$ for V1154~Cyg. {\it Bottom panel:} Fourier spectrum of the variation of the main pulsational frequency $(f_1(t))$ calculated from the analytical function. Both shows the modulation signal at $f=0.006327~{\rm d^{-1}}$. The temporal resolution of the latter method is limited to $< 0.02$ d$^{-1}$.}
\end{figure}

All the examined Fourier quantities and the $A_{1}$ amplitude show cycle-to-cycle variations as it was shown by \citet{der12}. In addition, we found periodic variations of the amplitudes and each Fourier quantity. We used {\sc Period04} by \citet{len05} and determined that period of the variation is  $ P_{\rm mod}=158.1 \pm 0.7$~d for $R_{21}$, $P_{\rm mod}=158.1 \pm 1.2$~d for $R_{31}$, while $P_{\rm mod}=159.9\pm 2.3 $~d for $\phi_{21}$ and $P_{\rm mod}=158.1 \pm 2.1 $~d for $\phi_{31}$. The period of the amplitude variation is slightly shorter,  $P_{\rm mod}=157.3 \pm 1.6 $~d. The differences likely arise from the presence of strong period jitter in the data. We conclude that there is a periodic modulation with an average period of $\sim 159$~d which was also detected in the frequency analysis (Sect.\ \ref{fouanal}). Our findings agree with the period calculated from smoothed Fourier parameters by \citet{kanev}. The frequency spectrum of $R_{21}$ is shown in the top panel of Fig.\ \ref{r21fou}.

We also used an alternative technique to determine the time dependence of the Fourier parameters, the analytical function method \citep{gabor}. The calculation is made in the Fourier space applying a filtering window around the examined frequency, and then fitting an assumed complex function in the form of $m(t)= A(t)\,e^{i \phi(t)}$ \citep{kollath01,kollath02}. This method provides instantaneous $A(t)$ amplitudes and $\phi(t)$ phases for the main pulsation frequency and its harmonics. The results of the analytical function analysis were identical to the ones mentioned above, confirming that the modulation exists in the data set, as the frequency spectrum of the $f_1(t)={\rm d}\phi_1 (t)/{\rm d}t$ parameter in the lower panel of Fig.\ \ref{r21fou} shows. We note that the width of the filtering window acts as a low-pass filter and limits the temporal resolution of the method: in this case the cutoff appears above 0.02~d$^{-1}$.

\section{The O--C diagram}
\label{oc}

Our previous papers \citep{sza11,der12} have already revealed short time-scale (even cycle-to-cycle) fluctuations in the pulsation period. Those papers were based on only partial {\it Kepler} data covering one and two years, respectively. As the {\it Kepler} space telescope continued collecting data until May 2013, the whole sample on V1154~Cygni covers four years. In addition to the LC data, V1154~Cyg was also observed in SC mode in 8 quarters, resulting in a superb coverage of photometric variations. The SC data of this Cepheid have not been studied before.

The analysis of the whole data set can be instrumental in confirming the strange behaviour revealed in this classical Cepheid and can lead to the discovery of additional peculiar phenomena unobservable in the previously studied shorter data segments.

The (in)stability of the pulsation period was studied by the method of the O--C diagram described in \citet{der12}. However, instead of constructing multiple O--C diagrams for various light curve phases, here we only studied temporal behaviour of the median brightness on the ascending branch whose moment can be determined much more accurately than that of other features on the light curve (see \citealt{der12}).

\begin{figure}
\begin{center}
\includegraphics[width=8cm]{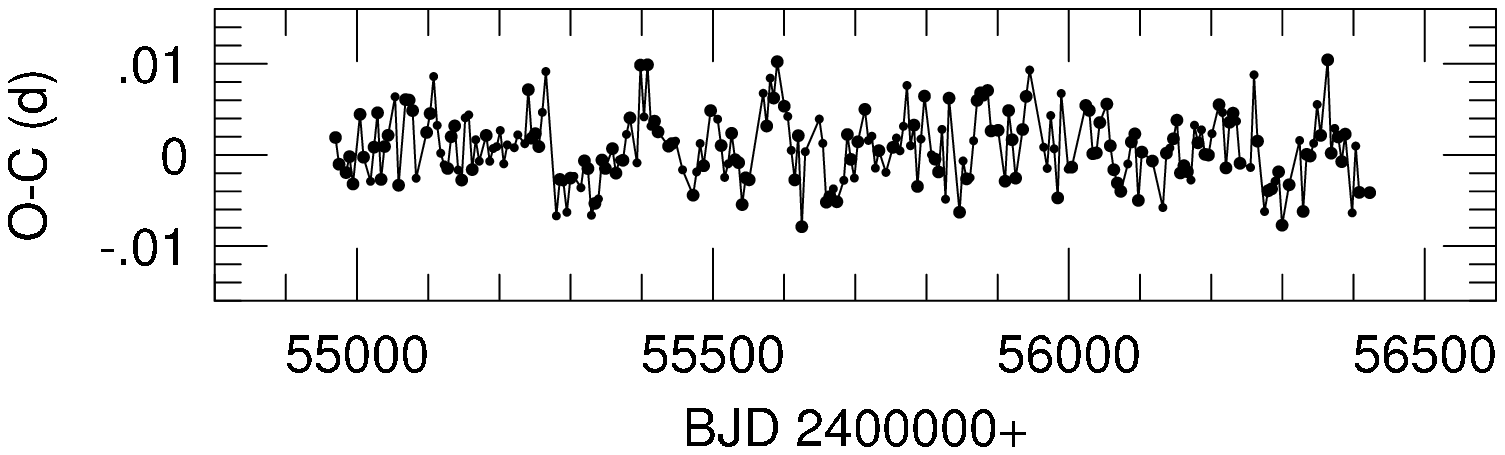}
\vskip5mm
\includegraphics[width=8cm]{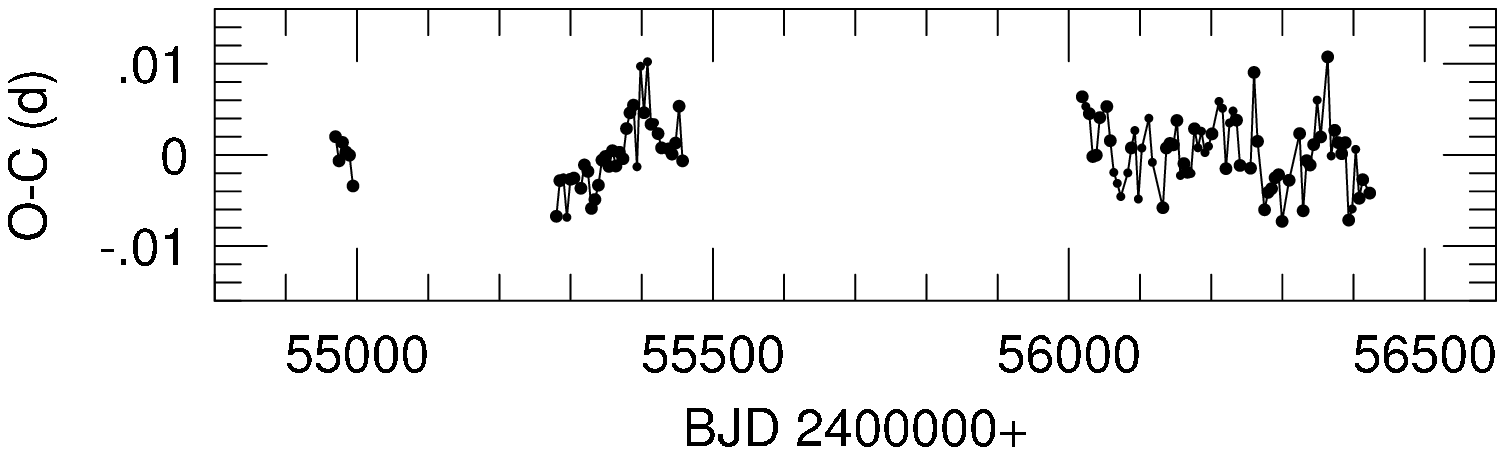}
\end{center}
\caption{\label{ocfig} {\it Top panel:} The O--C diagram of V1154~Cyg based on long cadence observations. {\it Bottom panel:} The O--C diagram of V1154~Cyg based on short cadence observations.}
\end{figure}

O--C diagrams have been constructed for both LC and SC data  (top panel and bottom panel of Fig.\ \ref{ocfig}, respectively, see also Tables\ \ref{tab-oc-lc}--\ref{tab-oc-sc}). The O--C differences for individual moments of median brightness on the ascending branch have been calculated by using the ephemeris:
${\rm C = BJD}\, (2454969.73958 \pm .00058)  + ( 4.925454 \pm 3\cdot10^{-6})\, {\rm E}$.
 
This ephemeris has been obtained by a weighted linear least squares fitting to the original O--C differences calculated by an arbitrary ephemeris. The scheme of weighting was very simple: normally a weight of unity is assigned to an O--C value. If, however, the light curve has not been perfectly covered near the brightness minimum and/or the subsequent brightness maximum, half weight is assigned to the relevant O--C difference. A jump exceeding one in the epoch numbering means that data are missing near minimum, median, or maximum brightness of the given (intermediate) pulsation cycle(s).

An immediate important result is related to the period value obtained by the weighted linear least squares fit. The average value of the pulsation period during the whole 4-year interval, 4.925454~d, is identical with the one obtained from the first year data in \citet{der12}. This points to the long-term stability of the pulsation period. 

\begin{table}
\begin{center}
\caption{O--C values of V1154~Cygni from the {\it Kepler} LC data 
(see the description in Sect.~\ref{oc}). A colon after the BJD
value means uncertain data. 
This is only a portion of the full version available online only.}
\label{tab-oc-lc}
\begin{tabular}{l@{\hskip2mm}r@{\hskip2mm}r}
\hline
\noalign{\vskip 0.2mm}
BJD$_{\odot}$ & $E\ $ & O--C\\
2\,400\,000 + & &\\
\noalign{\vskip 0.2mm}
\hline
\noalign{\vskip 0.2mm}
54969.74149 & 0 &    0.00191 \\
54974.66401 & 1 & $-$0.00102 \\
54984.51400 & 3 & $-$0.00194 \\
54989.44121 & 4 & $-$0.00019 \\
54994.36366 & 5 & $-$0.00319 \\
 \multicolumn{3}{l}{\dots}\\
\noalign{\vskip 0.2mm}
\hline
\end{tabular}
\end{center}
\end{table}

\begin{table}
\begin{center}
\caption{O--C values of V1154~Cygni from the {\it Kepler} SC data 
(see the description in Sect.~\ref{oc}). A colon after the BJD
value means uncertain data. 
This is only a portion of the full version available online only.}
\label{tab-oc-sc}
\begin{tabular}{l@{\hskip2mm}r@{\hskip2mm}r}
\hline
\noalign{\vskip 0.2mm}
JD$_{\odot}$ & $E\ $ & O--C \\
2\,400\,000 + & &\\
\noalign{\vskip 0.2mm}
\hline
\noalign{\vskip 0.2mm}
54969.74158 & 0 &    0.00200 \\
54974.66438 & 1 & $-$0.00065 \\
54979.59186 & 2 &    0.00137 \\
54984.51626 & 3 &    0.00032 \\
54989.44137 & 4 & $-$0.00003 \\
 \multicolumn{3}{l}{\dots}\\
\noalign{\vskip 0.2mm}
\hline
\end{tabular}
\end{center}
\end{table}

Study of the short time scale behaviour of the periodicity indicates that the instability of the pulsation from cycle to cycle keeps continuing. The dispersion of the O--C differences is about half an hour (0.004~$P$) in both O--C diagrams (Fig.\ \ref{ocfig}).

The comparison of the O--C diagrams for LC and SC data (top panel and bottom panel of Fig.\ \ref{ocfig}, respectively) shows that the scatter is real: the same patterns can be revealed in both diagrams (see especially the interval between BJD 2455300 and BJD 2455500). Although the SC and LC data are not independent as both are summed from the same individual exposures, the sampling frequency of the SC light curve is much higher. The O--C diagram for SC data is generally more accurate than its counterpart for the LC data, as beating between the sampling and the pulsation period decreases. Nevertheless, the dispersion of the O--C differences is practically the same in both diagrams. This fact also refers to reality of the period jitter.

\section{New radial velocities}
\label{newrv}

We obtained new spectroscopic observations in order to investigate the spectroscopic binary nature of V1154~Cyg. We took spectra with the new ACE spectrograph installed in 2014 on the 1~m RCC telescope at Piszk\'estet\H{o} Observatory (PO), Hungary. It is a fiber-fed spectrograph observing in the range of  4150--9150~\AA\ with a resolution of 20~000. Thorium-Argon (Th-Ar) lamp is available for the accurate wavelength calibration.

\begin{table}   
\begin{center}
 \caption{\label{rvtable} Log of the spectroscopic observations of V1154 Cyg containing date of observations, the computed RVs and their standard errors. The total uncertainty of the RVs, considering all the random and systematic errors, is estimated to be 0.5 km s$^{-1}$.}  
\label{maxtimes}  
\begin{tabular}{cr}  
\hline   
HJD & \multicolumn{1}{c}{RV} \\ 
(2 400 000+) & (${\rm km~s^{-1}}$) \\
\hline  
57099.59097	&	$-$8.10(13)	\\
57101.55819	&	7.69(15)	\\
57101.57916	&	7.61(14)	\\
57102.55296	&	$-$5.18(13)	\\
57102.57391	&	$-$5.77(13)	\\
57105.52855	&	1.09(13)	\\
57105.54953	&	1.11(14)	\\
57185.37562	&	8.46(16)	\\
57186.54199	&	$-$10.79(14)	\\
\hline   
\end{tabular} 
\end{center}  
\end{table}

We took nine spectra of V1154~Cyg on six nights in March and June 2015. The average $S/N$ of the spectra is $\sim$30. All spectra were reduced using IRAF\footnote{IRAF is distributed by the National Optical Astronomy Observatories, which are operated by the Association of Universities for Research in Astronomy, Inc., under cooperative agreement with the National Science Foundation.} standard tasks including bias and flat field corrections, aperture extraction, wavelength calibration (using Th-Ar lines) and barycentric correction. We then normalised the continuum of the spectra. RVs were determined by the cross-correlation method using metallic lines in the region between 4800 and 5600~\AA. We fitted the cross-correlated line profiles with a Gaussian function to determine the RVs, listed in Table~\ref{rvtable}.

Since these are the first published RV data obtained with this new instrument, here we briefly review the performance of the telescope and spectrograph system. We estimate the accuracy and precision of the measurements using observations of the RV standard $\beta$~CVn ($V_\mathrm{rad} = 6.259$\,km\,s$^{-1}$, \cite{nidever2002}) obtained each night when V1154~Cyg was observed. The $S/N$ of the $\beta$~CVn spectra is typically $\sim$60.

\begin{figure}
\begin{center}
\includegraphics[width=8cm]{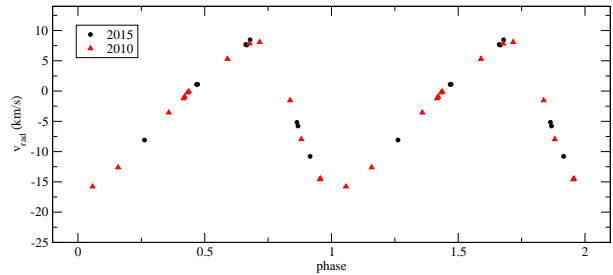}
\end{center}
\caption{\label{rv} Phase folded RV data of V1154~Cyg using the period of 4.925454~d from \citet{der12}. The black circles denote the new data obtained in 2015 compared to the data published by \citet{der12} plotted with red triangles. The RV data do not show any $\gamma$-velocity change.}
\end{figure}

\begin{table} 
\begin{center}
 \caption{\label{tbl:rvstd} Log of the spectroscopic observations of the RV standard $\beta$~CVn containing date of observations, the computed RVs and their standard errors.}
\begin{tabular}{cr}
\hline 
HJD & \multicolumn{1}{c}{RV} \\
(2 400 000+) & (${\rm km~s^{-1}}$) \\
\hline
 57099.39360  &  6.40(15)  \\
 57101.43988  &  6.91(16) \\
 57102.43448  &  6.03(15) \\
57105.44771  & 5.99(15) \\
57185.34181  & 6.67(15) \\
57186.32281  & 6.46(17) \\

\hline
\end{tabular}
\end{center}
\end{table}

We processed the RV standard observations the same way as described for V1154~Cyg. The obtained radial velocities of $\beta$~CVn are listed in Table~\ref{tbl:rvstd}. The errors given in Tables~\ref{rvtable} and \ref{tbl:rvstd} correspond to the standard zero-point error of the fitted Gaussian function. The mean and standard deviation of the measurements is $6.41\pm0.36$\,km\,s$^{-1}$, in agreement with the literature value. The standard deviation is twice larger than the precision of the individual observations. This is explained by the systematic errors introduced by the ThAr RV calibration, normalisation uncertainties and the choice of the line mask for the cross-correlation. The bias introduced by the latter two was estimated by performing the normalisation in several different ways, and calculating the cross-correlation with different sub-sets of the full line list. The effects of the choice of normalisation and line mask are 0.03\,km\,s$^{-1}$ and 0.2\,km\,s$^{-1}$, respectively. We conclude that the wavelength calibration system does not have a systematic zero-point error larger than 0.36\,km\,s$^{-1}$.

Due to the significant temperature variations, the spectral type of V1154~Cyg changes somewhat during the pulsation, which introduces further pulsation-phase dependent systematic errors, similarly to those caused by the line mask choice. Therefore, we estimate the total systematic and random errors of the RVs of V1154~Cyg to be 0.5~${\rm km~s^{-1}}$.

\section{Discussion}
\label{discussion}

\subsection{Any secondary component?}

We checked whether the $\sim$159~d modulation is caused by the presence of a companion star. Assuming that the  $\sim$159~d modulation seen in the O--C diagram is caused by the light-time effect, i.e. to orbital motion in a binary system, and circular orbit, we can calculate the RV amplitude of the Cepheid variable.

The semi-amplitude of the O--C in Fig.\ \ref{ocfig} is $A_{\rm (O-C)}= 0.0038$~d, so the expected RV amplitude is $\sim$45~km ${\rm s^{-1}}$. This $\gamma$-velocity variation is clearly not seen in Fig.\ \ref{rv}, nor in Fig.~14 in \citet{sza11}, where all the RV data are collected from the literature.

Therefore, we can conclude that the detected $\sim$159~d modulation is not due to the presence of a companion star.

\subsection{Blazhko-like modulation(s) of V1154\,Cygni }

As we mentioned previously, the $\sim$ 159-d modulation was detected both in the Fourier parameters and in the O--C variations. The variation of the Fourier parameters corroborates our previous finding of light curve shape variations in this Cepheid. We note that the existence of light curve shape variation resembles the one seen in RR~Lyrae stars caused by the Blazhko effect (e.g. {\it Kepler}: \citet{kol11}; {\it CoRoT}: \citet{gug11}), and the simultaneous, cyclic frequency variation and light curve shape deformation also points to a Blazhko-like mechanism.

Although Blazhko-like variations are very rare among single-mode Cepheids, there are a few such objects that may display similar modulation. The most prominent example within the Milky Way is V473 Lyr \citep{mol13,mol14}, which pulsates in the second overtone and has also two distinct modulation periods, 1204 d and 14.5 yr. Another well-known example is Polaris ($\alpha$ UMi) that went through a low-amplitude phase during the last few decades \citep{bru08}. However, even if the variations in Polaris eventually turn out to be periodic, they would represent an extreme case of centuries-long modulation. Recently, a few more promising candidates have been found, such as SV Vul that shows a $\sim 30$ year cycle in its pulsation period and possibly in its amplitude too \citep{engle2015}. However, these stars have cycle lengths in the range of decades and will require further observations to determine the exact nature of period and amplitude changes. A slightly different candidate is $\ell$ Car where changes in the RV amplitude were observed over the span of only two years, suggesting a shorter timescale, similar to V1154~Cyg, although no cycle length was determined for the star itself \citep{anderson2014}. 

We also note that several single-mode Cepheids within the Magellanic Clouds display periodic amplitude variations. However, most of them show only a single frequency peak next to the pulsation frequency, instead of symmetrical triplets, suggesting that in those cases the modulation is likely caused just by the beating between two closely-spaced pulsation modes \citep{soszynski2008,moskalikkov2009}. Recently, \citet{soszynski2016} described three Cepheids in the Large Magellanic Cloud that appear to show proper Blazhko-effect akin to that of V473~Lyrae, with modulation periods in the 1000--5000 d range.

Whether these stars are modulated by the same mechanism as RR~Lyrae stars is not known yet. It also remains open whether the strongly modulated, second-overtone star V473~Lyr, and V1154~Cyg, which pulsates in the fundamental radial mode, share the same physical explanation of their variation. Nevertheless, the phenomenological similarity points towards a common mechanism.

\subsection{The granulation noise}

\begin{figure}
\includegraphics[width=1.0\columnwidth]{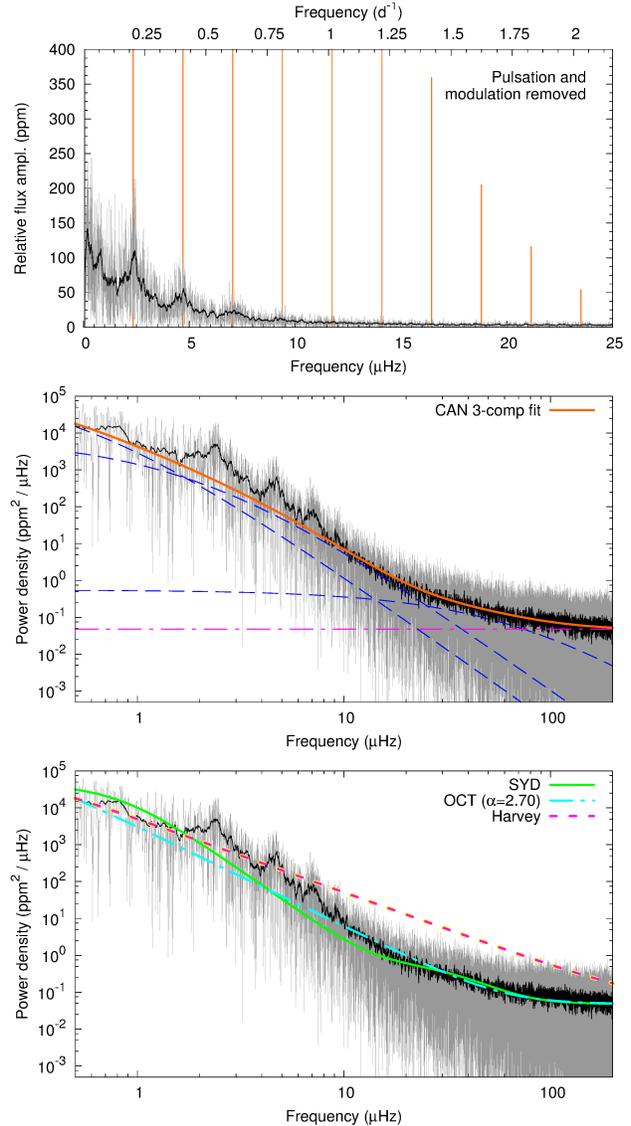}
\caption{Top: the Fourier spectrum of the residual data after we prewhitened with 28 pulsation and modulation frequency components and filled the gaps. The positions of the $nf_1$ components are indicated with the orange lines. The full-resolution spectrum is shown in grey, a smoothed version in black. Middle: the same, converted to power density spectrum. The solid orange line shows the fitted granulation signal, comprising of three super-Lorentzian functions (dashed blue lines), plus a constant noise component (the dash-dotted line), following the CAN method. Bottom: the granulation signal, fitted with three other methods, with varying degrees of success. }
\label{granulation}
\end{figure}

After prewhitening with the pulsation and modulation peaks, we are left with a Fourier spectrum that still contains significant power distributed into a forest of peaks between 0.0 and about 1.0 d$^{-1}$ frequency interval, increasing towards zero frequency (Figures \ref{granulation} and \ref{lorentz}). Some part of this residual power likely comes from the inaccuracies of stitching the {\it Kepler} quarters together, but the experience with RR~Lyrae light curves suggests that instrumental signals mostly manifest as few discrete peaks related to the length of the quarters and the orbital period of \textit{Kepler}, especially if the light curve contains quarter-long gaps \citep{ben14}. In this case, however, we see a combination of  a wider, red noise-like component ($\sim f^{-2}$) with additional increases around the positions of the $nf_1$ frequency peaks. 

In stars that show solar-like oscillations, red noise is usually attributed to granulation, the overall effect of many convective cells appearing and disappearing in the photosphere. For the smaller red giants, up to about 25 R$_\odot$, we are now able to disentangle solar-like oscillations and granulation on a regular basis with data from photometric space telescopes like \textit{Kepler} (see, e.g., \citealt{mathur2011,kallinger2014}). For the largest, nearby red supergiants, such as Betelgeuse ($\alpha$ Ori) we can obtain direct observations via disk-resolved imaging or interferometry that show actual hot spots on the surface, and compare those to model calculations \citep{chiavassa2010}. Cepheids also possess envelope convection zones, therefore it is reasonable to expect that some form of granulation can occur in them as well. Furthermore, they lie between the oscillating red giants and the red supergiants in terms of size and mass, so comparison with both groups may be of importance. Visible-light observations obtained with the \textit{WIRE} space telescope suggested a small low-frequency excess in Polaris that was presumed to originate from granulation \citep{bru08}. However, both the frequency resolution and the precision of those data were much lower than that of V1154 Cyg. 

We computed the power density spectrum of the star after removing the pulsation and modulation signals from the light curve. \citet{kallingermatthews2010} followed a similar procedure for $\delta$ Scuti stars: first they prewhitened the data with 10 strongest frequency components, and then showed that the residual frequency spectra can be adequately described as a sum of granulation noise and a moderate number of pulsation modes. In the case of V1154~Cyg, we removed 28 frequency components and continued the analysis of the residual light curve. We filled the gaps in the residual light curve with linear interpolations, following the recipe of \citet{kallinger2014} to avoid leaking excess amplitude into the high-frequency regime through the wing structures of the spectral window. In addition, we calculated similar power density spectra for a few RR Lyrae stars as comparison, after prewhitening with the main harmonics and the modulation triplet peaks ($f_n\pm f_m$). The power density spectra of the RR Lyrae stars turned out to be flat, lacking any signs of excess noise that may arise either from granulation or instrumental effects, indicating that the signal in V1154 Cyg is intrinsic to the star.

The granulation noise was first modeled for the Sun by \citet{harvey1985}. He concluded that the autocovariance of the granulation velocity field follows an exponential decay function, leading to a Lorentzian profile in the power density spectrum of the measured variations (note that the original notation there uses $\nu$ for frequency instead of $f$):

\begin{equation} P(f) \sim \frac{\zeta\,\sigma^2\tau_{\rm gran}}{1+(2\pi\, f\, \tau_{\rm gran})^\alpha} \end{equation}

\noindent where $\tau_{\rm gran}$ is the characteristic time scale and $\sigma$ is the characteristic amplitude of the granulation. The numerator, $P_{\rm gran} = \zeta\,\sigma \tau_{\rm gran}^2$ is defined as the amplitude of granulation power, with $\zeta$ being a normalisation factor that depends on the value of $\alpha$ \citep{kallinger2014}. For $\alpha = 2$ and  $\alpha=4$, $\zeta$ can be analytically calculated to be 4 and $4\sqrt{2}$, respectively. \citet{harvey1985} originally employed $\alpha = 2$, i.e., a true Lorentzian function. Since then, several methods have been developed to investigate the granulation background of other stars (e.g., CAN: \citealt{kallinger2010}; OCT: \citealt{hekker2010}; SYD: \citealt{huber2009}). An important modification is that nowadays most methods use different exponents in the denominator that are usually higher than 2, called `super-Lorentzian' functions. We tested various prescriptions based on the comparative study of  \citet{mathur2011}. The results are summarised in Figure \ref{granulation}: as the bottom panel shows, the original, single-component, $\alpha=2$ function of \citet{harvey1985}, and the SYD method (a combination of $\alpha=2$ and 4 functions) does not fit the data well. If we use the OCT method and use $\alpha$ as a free parameter, we find a fairly good fit at $\alpha=2.70$.  The CAN method (top panel) that combines three $\alpha=4$ functions fits the data well too. In all cases the region with excess amplitude, between 2--8 $\mu$Hz (0.17-0.69 d$^{-1}$) has been excluded from the fit. 

With the invention of different methods, the comparison of various $\tau$ values became problematic. In the case of the CAN method, for example, $\tau_2 = \tau_{\rm gran}$, but the other two $\tau$ parameters have less well defined physical meanings. To solve this issue, \citet{mathur2011} proposed to calculate an effective timescale ($\tau_{\rm eff}$) instead, the e-folding timescale of the autocorrelation function. In the case of V1154~Cyg, we found that $\tau_{\rm eff} = (3.0  \pm 0.4)\, 10^5$ s, or  $3.5 \pm 0.5$ d, in agreement with the $\tau_2$ parameter of the CAN method. We calculated the granulation power, defined as $P_{\rm gran} = \sum \zeta\,\sigma_i^2\tau_i$, and the amplitude of the intensity fluctuation, $A_{\rm gran}^2 = C_{\rm bol}^2 \sum (\sigma_i^2 / \sqrt{2})$ (where  $C_{\rm bol}^2$ is a bolometric correction factor, as defined by \citealt{ballot2011}), to be $P_{\rm gran} = (2.65\pm 0.37)\, 10^5$ ppm$^2/\mu$Hz, and $A_{\rm gran} = 256 \pm 34$ ppm, respectively \citep{mathur2011,kallinger2014}.  The values are summarised in Table \ref{table:gran}.

\subsection{Granulation properties}

\begin{figure}
\includegraphics[width=1.0\columnwidth]{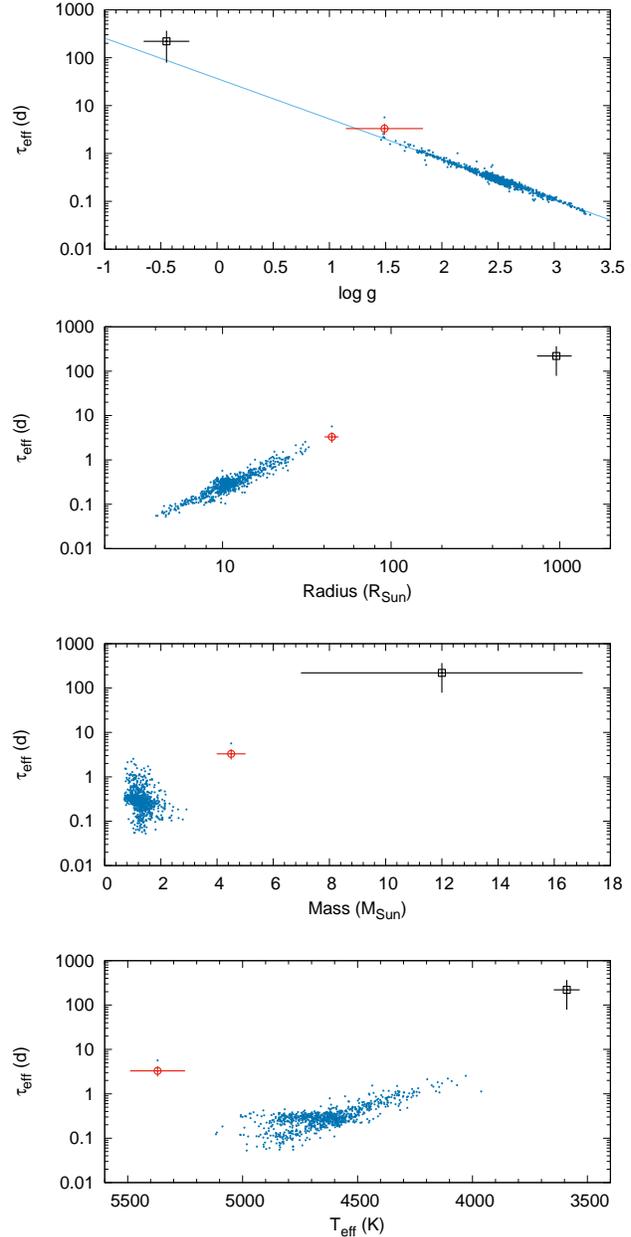}
\caption{Comparison of the $\tau_{\rm eff}$ effective timescale granulation parameter of V1154~Cyg (red circle), Betelgeuse (black square) and the \textit{Kepler} red giant sample of \citet{mathur2011}. Values are plotted against the stellar radius, logarithm of surface gravity, mass and effective temperature, respectively. The blue line in the top plot is an extrapolation of the scaling based on the red giant data. }
\label{gran_comp}
\end{figure}

To investigate the derived granulation parameters, we compared them to the red giant sample presented by \citet{mathur2011} and the well-observed supergiant, Betelgeuse (Figure \ref{gran_comp}). For the latter star, we analyzed the visual light curve data from the AAVSO (American Association of Variable Star Observers) database. The power density spectrum of Betelgeuse exhibits an obvious slope that can be fitted with various granulation noise curves. The light variations of the star are complicated by the semiregular variations, so we did not attempt a detailed analysis in this paper, instead we estimated $\tau_{\rm eff}$ only. Different fits to the autocorrelation function resulted in values between 120 and 435 d. We settled for a value of $\tau_{\rm eff} \approx 280 \pm 160$ d. Despite the large uncertainty, this confirms that the granulation timescale in red supergiants is in the range of a year. It is also in agreement with the 400-day timescale of convective motions in Betelgeuse, derived from line bisector variations \citep{gray2008}. Mass, radius, and log\textit{g} data for Betelgeuse are from the studies of \citet{lobeldupree2000} and \citet{neilson2011}. 

\begin{table}
\begin{center}
\caption{Granulation parameters for V1154~Cyg.} 
\begin{tabular}{ccc}
\hline
$\tau_{\rm eff}$   &    $P_{\rm gran} $        &   $A_{\rm gran}$ \\
(sec)                   &      (ppm$^2/\mu$Hz)   &   (ppm)                \\
\noalign{\vskip 0.1cm}    
\hline
$(3.0  \pm 0.4)\, 10^5$   &  $(2.65\pm 0.37)\, 10^5$   &   $256 \pm 34 $  \\
\hline
\end{tabular}
\label{table:gran}
\end{center}
\end{table}

Figure \ref{gran_comp} shows that the scaling between the derived effective granulation timescale, $\tau_{\rm eff}$ and log $g$ is acceptable for both stars: they fall close but somewhat above the extrapolated power law based on the red giant data. Similar, but less tight scaling can be observed with the radius and mass as well. The effective temperature plot, on the other hand, is dominated by stellar evolution and it reproduces the positions of the stars in the Hertzsprung-Russell diagram instead. 

For V1154~Cyg we have a reliable estimate for the granulation power ($P_{\rm gran}$) too. The ratio of the two quantities, $P_{\rm gran}/\tau_{\rm eff}$, is related to the variance of the intensity variations over the stellar surface. This ratio is $\approx 0.9$ for the Cepheid but it is $\sim 10$ for red giants with similar log $g$ values (see Figure 5 by \citealt{mathur2011}). The difference can be attributed to various factors, e.g. lower contrast between cool and dark regions, or smaller cell sizes, leading to stronger cancellation effect over the surface. 

\subsection{Non-detection of solar-like oscillations}

The observations of \textit{Kepler} revealed that solar-like oscillations in red giant stars have larger amplitudes and lower characteristic frequencies as the mass and size of the stars increase \citep{huber2011}. At first we searched for regular frequency spacings, a characteristic feature of the stochastically-driven oscillations, in the residual spectrum, but we found no significant patterns. Then we calculated the approximate range of $f_{\rm max}$ (or $\nu_{\rm max}$, in the notation used for solar-like oscillations), the frequency and amplitude of maximum oscillation power for V1154~Cyg, assuming a mass range of 4--8 M$_\odot$, based on the scaling relations of \citet{huber2011}. For this Cepheid, $f_{\rm max}$ would be between 5--10 $\mu$Hz, with an associated amplitude increasing from a minimum of 80 ppm to roughly 500 ppm, as the assumed $f_{\rm max}$ decreases. However, after we removed the granulation noise from the residual Fourier spectrum of V1154~Cyg, the remaining signals are about a factor of 4--5 smaller than the expected oscillation amplitudes in the entire frequency range. The only discernible features between 5--10 $\mu$Hz are the small excesses around the positions of the 3$f_1$ and 4$f_1$ frequency peaks. This result suggests that large-amplitude, coherent stellar pulsation suppresses or completely blocks solar like oscillations in Cepheids. Inhibition of regular oscillations was also observed in compact triple-star systems where tidally induced oscillations are excited instead \citep{derekas2011,fuller2013}.

\begin{figure}
\includegraphics[width=1.0\columnwidth]{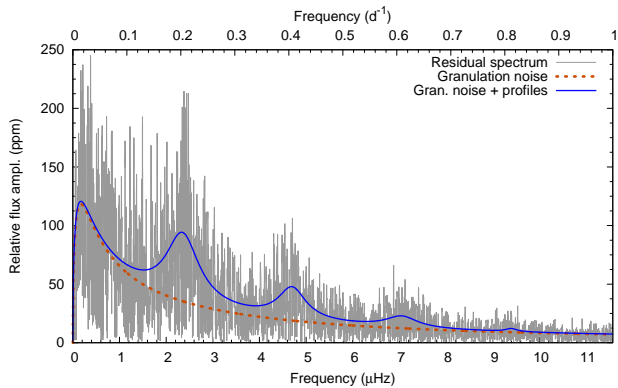}
\caption{The residual Fourier spectrum of V1154~Cyg with two different fits: the orange dashed line is the granulation noise profile only, the blue line is granulation plus Lorentzian profiles at the broad excesses. }
\label{lorentz}
\end{figure}

The remaining signals in the Fourier-spectrum are the broad forests around the positions of the pulsation frequencies. Forests of peaks indicate a non-coherent signal, therefore we fitted the positions of the first four harmonics from $f_1$ to $3f_1$, with Lorentzian profiles added to the granulation noise (Figure~\ref{lorentz}). The full width at half maximum values of the profiles indicates that the lifetime is $\tau = 1/\pi\Gamma = 4.5 \pm 0.4$ d. The origin of these profiles is not completely clear: they could originate from the pulsation jitter, but the Eddington-Plakidis test conducted in the previous study of V1154~Cyg indicate that the fluctuations average out on a longer time scale of 15 d \citep{der12}. Another possibility is that the signal originates from the interaction between granulation (e.g., convective motions) and stellar pulsation. 

\section{Summary}
\label{summary}

In this paper, we presented the analysis of the {\it Kepler} field Cepheid V1154~Cyg using the whole available long and short cadence {\it Kepler} photometry spanning 1460~d (omitting data from Q0). For the analysis, we applied aperture photometry to the {\it Kepler} pixel data. The main results are summarised as follows:
\begin{itemize}

\item The Fourier analysis yielded 28 frequencies. The main pulsation frequency is at $f_{1}=0.203027$~c/d. Besides identifying the harmonics of $f_{1}$, we also detected the modulations of $f_{1}$, $f_{2}$, $f_{3}$  with a period of $\sim$159~d and $\sim$1160~d, respectively.

\item In order to study the light curve shape we calculated the Fourier parameters of each pulsational cycle.  We concluded that there is a periodic variations of the amplitude and each Fourier quantity with an average period of $\sim$159~d.

\item Besides the short term fluctuation of the O--C diagram (which was first detected in \citet{der12}), we detected some periodic variations with $\sim$159~d.

\item We identified the residual power at low frequencies as granulation noise, and determined its effective timescale to be $\tau_{\rm eff} = 3.5 \pm 0.5$ d. The timescale agrees with the scalings with fundamental physical parameters observed in red giants quite well.

\item We did not detect any signs of solar-like oscillations, down to levels 4--5 times smaller than the expected oscillation amplitudes, suggesting that coherent pulsations suppress oscillations in Cepheids.

\item We obtained new radial velocity data and compared them with previous data from \citet{der12} and did not find any $\gamma$-velocity change. The radial velocities also revealed that the $\sim$159~d modulation is not caused by the presence of a companion star.

\end{itemize}

The findings presented in this paper offer tantalising clues about the intimate connection between pulsation and convection in Cepheids. The presence of modulation, period jitter, and granulation, and the apparent lack of evanescent oscillations indicate that several questions still surround the internal behaviour of these stars. The maturation of multi-dimensional hydrodynamic models promise that theoretical studies will soon be able to tackle these problems \citep{mundprecht2015}. From the observational side, further examples for these effects will require long, continuous, high-precision measurements that space-based missions can provide. The {\it TESS} mission will collect data for about one year in the continuous viewing zones (CVZ) around the Ecliptic Poles \citep{ricker2014}. The Southern CVZ contains one bright Cepheid, $\beta$ Dor that could be subject of similar studies. However, we likely have to wait until the start of {\it PLATO} \citep{rauer2014} to target more Cepheids and to match the unique capabilities the original \textit{Kepler} mission provided.

\section*{Acknowledgments} 

Funding for the \textit{Kepler} and K2 missions is provided by the NASA Science Mission directorate. We thank Savitha Mathur for providing us with the red giant granulation parameters. This project has been supported by the Hungarian NKFI Grants K-113117, K-115709, K-119517, PD-116175 and PD-121203 of the Hungarian National Research, Development and Innovation Office, the ESA PECS Contract No. 4000110889/14/NL/NDe, the Lend\"ulet-2009, LP2012-31 and LP2014-17 programs of the Hungarian Academy of Sciences. AD has been supported by the Postdoctoral Fellowship Programme of the Hungarian Academy of Sciences and by the Tempus K\"ozalap\'itv\'any and the M\'AE\"O. LM and \'AS have been supported by the J\'anos Bolyai Research Scholarship of the Hungarian Academy of Sciences. AD, BCs and GyMSz would like to thank the City of Szombathely for support under Agreement No. 67.177-21/2016. We acknowledge with thanks the variable star observations from the AAVSO International Database contributed by observers worldwide and used in this research. The authors gratefully acknowledge the Kepler Science Team and all those who have contributed to the Kepler Mission for their tireless efforts which have made these results possible.


\begin{thebibliography}{}

\bibitem[\protect\citeauthoryear{Anderson}{2014}]{anderson2014} Anderson R.\ I., 2014, IAUS, 307, 286

\bibitem[Arellano Ferro et~al.(1998)]{are98}
 Arellano Ferro A., Rojo Arellano R., Gonz\'alez-Bedolla S., Rosenzweig P., 1998, ApJS, 117, 167

\bibitem[\protect\citeauthoryear{Ballot et~al.}{2011}]{ballot2011} 
 Ballot J., Barban C., Van't Verr-Menneret C., 2011, A\&A, 531, 124

\bibitem[Benk\H{o} et~al.(2014)]{ben14}
 Benk\H o J. M., Plachy E., Szab\'o R., Moln\'ar L., Koll\'ath Z., 2014, ApJS, 213, 31

\bibitem[Benk\H{o} \& Szab\'o(2015)]{ben15}
 Benk\H o J. M.,  Szab\'o R., 2015, ApJ, 809, 19

\bibitem[Berdnikov(2008)]{ber08}
 Berdnikov L. N., 2008, VizieR On-line Data Catalog: II/285

\bibitem[Breger et~al.(1993)]{bre93}
 Breger M., et~al., 1993, A\&A, 271, 482

\bibitem[Bruntt et~al.(2008)]{bru08}
 Bruntt H., et~al., 2008, ApJ, 683, 433

\bibitem[\protect\citeauthoryear{Chiavassa et~al.}{2010}]{chiavassa2010} 
 Chiavassa A., Haubois X., Young J. S., Plez, B. Josselin E., Perrin G., Freytag B., 2010, A\&A, 515, 12

\bibitem[\protect\citeauthoryear{Derekas et~al.}{2011}]{derekas2011} 
 Derekas A., et~al., 2011, Science, 332, 216

\bibitem[Derekas et~al.(2012)]{der12}
 Derekas A., et~al., 2012, MNRAS, 425, 1312

\bibitem[\protect\citeauthoryear{Engle}{2015}]{engle2015} 
Engle S.\ G., 2015, PhD Thesis, James Cook University, arXiv:1504.02713

\bibitem[ESA(1997)]{esa97}
 ESA 1997, Hipparcos Catalogue, ESA SP-1200

\bibitem[Evans et~al.(2015)]{eva15}
 Evans N. R., et~al., 2015, MNRAS, 446, 4008 

\bibitem[\protect\citeauthoryear{Fuller et~al.}{2013}]{fuller2013} 
 Fuller J., Derekas A., Borkovits T., Huber D., Bedding T.\ R., Kiss L.\ L., 2013, MNRAS, 429, 2425

\bibitem[\protect\citeauthoryear{G\'abor}{1946}]{gabor} 
 G\'abor D., 1946, \textit{Theory of communication}, J. Inst. Electr. Eng. III, 93, 429

\bibitem[Gilliland et~al.(2010)]{gil10}
 Gilliland R. L., et~al., 2010, ApJL, 713, L160

\bibitem[Gorynya et~al.(1998)]{gor98}
 Gorynya N. A., Samus’ N. N., Sachkov M. E., Rastorguev A. S., Glushkova E. V., Antipin S. V., 1998, Astron. Lett., 24, 815

\bibitem[\protect\citeauthoryear{Gray}{2008}]{gray2008} 
 Gray D.\ F., 2008, AJ, 135, 4

\bibitem[Guggenberger et~al.(2011)]{gug11}
 Guggenberger E., Kolenberg K., Chapellier E., Poretti E., Szab\'o R., Benk\H o J. M., Papar\'o M., 2011, MNRAS, 415, 1577

\bibitem[\protect\citeauthoryear{Harvey}{1985}]{harvey1985} 
 Harvey J., 1985, in Future Missions in Solar, Heliospheric \& Space Plasma Physics, eds. E. Rolfe, \& B. Battrick, ESA SP-235, 199

\bibitem[\protect\citeauthoryear{Hekker et al.}{2010}]{hekker2010} 
Hekker S., et al., 2010, MNRAS, 402, 2049

\bibitem[\protect\citeauthoryear{Huber et~al.}{2009}]{huber2009} 
Huber D., et al., 2009, CoAst, 160, 74

\bibitem[\protect\citeauthoryear{Huber et~al.}{2011}]{huber2011} 
 Huber D., et~al., 2011, ApJ, 743, 143

\bibitem[Ignatova \& Vozyakova(2000)]{ign00}
 Ignatova V. V., Vozyakova O. V., 2000, Astron. Astrophys. Trans., 19, 133

\bibitem[Imbert(1999)]{imb99}
 Imbert M., 1999, A\&AS, 140, 79

\bibitem[Jenkins et~al.(2010a)]{jen10a}
 Jenkins  J.~M., et~al., 2010a, \apj, 713, L87

\bibitem[Jenkins et~al.(2010b)]{jen10b}
 Jenkins  J.~M., et~al., 2010b, \apj, 713, L120

\bibitem[Jerzykiewicz(2013)]{jer13}
 Jerzykiewicz M., 2013, Astrophys. Space Sci.\ Proc.\ Vol.\ 31.,  pp 291-293, Pulsating stars: Impact of New Instrumentation and New Insights, Poster No. 30.

\bibitem[\protect\citeauthoryear{Kallinger \& Matthews}{2010}]{kallingermatthews2010} 
 Kallinger T., Matthews J., 2010, ApJ, 711, L35

\bibitem[\protect\citeauthoryear{Kallinger et~al.}{2010}]{kallinger2010}  
 Kallinger, T., et al. 2010, A\&A, 522, A1

\bibitem[\protect\citeauthoryear{Kallinger et~al.}{2014}]{kallinger2014} 
 Kallinger T., et~al., 2014, A\&A, 570, A41

\bibitem[\protect\citeauthoryear{Kanev, Savanov \& Sachkov}{2015}]{kanev} 
 Kanev E., Savanov I., Sachkov M., 2015, EPJ WoC, 101, 06036 

\bibitem[Koch et~al.(2010)]{koc10}
 Koch  D.~G., et~al., 2010, \apj, 713, L79

\bibitem[Kolenberg et~al.(2011)]{kol11}
 Kolenberg K., et~al., 2011, MNRAS, 411, 878

\bibitem[\protect\citeauthoryear{Koll\'ath \& Buchler}{2001}]{kollath01} 
 Koll\'ath Z., Buchler J. R., 2001, in Astrophys.\ Space Sci.\ Libr.\ Ser.\ Vol. 257, Stellar Pulsation -- Nonlinear Studies. Kluwer, Dordrecht, p.\ 29

\bibitem[\protect\citeauthoryear{Koll\'ath et~al.}{2002}]{kollath02} 
 Koll\'ath Z., Buchler J. R., Szab\'o R., Csubry Z., 2002, A\&A, 385, 932

\bibitem[Lenz \& Breger(2005)]{len05}
 Lenz P., Breger, M., 2005, Commun. Asteroseismol., 146, 53

\bibitem[\protect\citeauthoryear{Lobel \& Dupree}{2000}]{lobeldupree2000} 
 Lobel A., Dupree A.\ K., 2000, ApJ, 545, 454

\bibitem[Luck, Kovtyukh \& Andrievsky(2006)]{luc06}
 Luck R. E., Kovtyukh V. V., Andrievsky S. M., 2006, AJ, 132, 902

\bibitem[\protect\citeauthoryear{Mathur et~al.}{2011}]{mathur2011} 
 Mathur S., et~al., 2011, ApJ, 741, 119

\bibitem[Molenda-\.Zakowicz, Frasca \& Latham(2008)]{mol08}
 Molenda-\.Zakowicz J., Frasca A., Latham D. W., 2008, Acta Astron., 58, 419

\bibitem[Moln\'ar et~al.(2013)]{mol13}
 Moln\'ar L., Szabados L., Dukes R. J., Jr., Gy\H{o}rffy \'A., Szab\'o R., 2013, AN, 334, 980 

\bibitem[Moln\'ar \& Szabados(2014)]{mol14}
 Moln\'ar L., Szabados L., 2014, MNRAS, 442, 3222

\bibitem[\protect\citeauthoryear{Moskalik \& Ko\l{}aczkowski}{2009}]{moskalikkov2009} 
 Moskalik P.\ A., Ko\l{}aczkowski Z., 2009, MNRAS, 394, 1649

\bibitem[\protect\citeauthoryear{Mundprecht, Muthsam \& Kupka}{2015}]{mundprecht2015} 
 Mundprecht E., Muthsam H.\ J., Kupka F., 2015, MNRAS, 449, 2539

\bibitem[\protect\citeauthoryear{Neilson, Lester \& Haubois}{2011}]{neilson2011} 
 Neilson H.\ R., Lester J.\ B., Haubois X., 2011, ASPC, 451, 117

\bibitem[\protect\citeauthoryear{Nidever et~al.}{2002}]{nidever2002} 
 Nidever D. L., Marcy G. W., Butler R. P., Fischer D. A., Vogt S. S. 2002, ApJS, 141, 503

\bibitem[Pigulski et~al.(2009)]{pig09}
 Pigulski A., Pojma\'nski G., Pilecki B., Szczygie{\l} D. M., 2009, Acta Astron., 59, 33
 
\bibitem[\protect\citeauthoryear{Rauer et~al.}{2014}]{rauer2014} 
 Rauer H., et~al., 2014, Exp.\ Astr., 38, 249

\bibitem[\protect\citeauthoryear{Ricker et~al.}{2014}]{ricker2014}  
 Ricker G. R., et~al., 2014, Proc. SPIE, 9143, 914320

\bibitem[Simon \& Lee(1981)]{sim81}
 Simon N. R., Lee A. S., 1981, ApJ, 248, 291 

\bibitem[\protect\citeauthoryear{Soszy\'nski et~al.}{2008}]{soszynski2008} 
 Soszy\'nski I., et~al., 2008, Acta Astr., 58, 163 

\bibitem[\protect\citeauthoryear{Soszy\'nski et~al.}{2015}]{soszynski2015} 
 Soszy\'nski I., et~al., 2015, Acta Astr., 65, 329 

\bibitem[Strohmeier, Knigge \& Ott(1963)]{str63}
 Strohmeier W., Knigge R., Ott H., 1963, Bamberg Ver\"off., V, Nr. 16

\bibitem[Szabados(1977)]{sza77}
 Szabados L., 1977, Commun. Konkoly. Obs. Hung. Acad. Sci., Budapest, No. 70

\bibitem[Szab\'o et~al.(2011)]{sza11}
 Szab\'o R., et~al., 2011, \mnras, 413, 2709
 
\bibitem[Wachmann(1976)]{wac76}
 Wachmann A. A., 1976, A\&AS, 23, 249

\end{thebibliography}
\end{document}